# Experimental demonstration of a fifth force due to chameleon field via cold atoms

Hai-Chao Zhang*

Key Laboratory for Quantum Optics, Shanghai Institute of Optics and Fine Mechanics, CAS, Shanghai 201800, China.

* E-mail: zhanghc@siom.ac.cn

**Abstract**: We tested a fifth force using cold atom experiments. The accelerated expansion of the universe implies the possibility of the presence of a scalar field throughout the universe driving the acceleration. This field would result in a detectable force between normal-matter objects. Theory of the chameleon field states that the force should be strong in a thin shell near the surface of a source object but greatly suppressed inside and outside of the source object. We used two atom clouds: one as the source and the other as the test mass; so the test mass can pass through the thin-shell region of the source mass. We detected the chameleon force and obtained the couple constant of about $4.5 \times 10^{11}$ between matter and the field. The chameleon force is considerably larger than Newtonian gravity at short distance; the interaction range is short enough to satisfy all experimental bounds on deviations from general relativity.

**One Sentence Summary:** A new force is detected by cold atom experiments and the couple constant between matter and chameleon field is estimated to be $4.5 \times 10^{11}$.

**Main Text:**

Astrophysical observations indicate that the expansion of the universe is accelerating (*1, 2*). This acceleration can be explained by dark energy component with negative pressure (*3*). The nature and origin of this energy are not understood. A possibility is that it originates from a new scalar field coupling to ordinary matter. If the scalar field exists, it should produce a new 'fifth force' according to quantum field theory (*4*). The most sensitive previous searches for violations of Newtonian gravity at submillimeter regime show that any fifth force is far weaker than gravity (*5*). Thus, two of the primary goals in this research area arise: 1. Scalar field theories for dark energy require a screening mechanism to explain why the fifth force has not yet been observed in laboratory and in solar-system experiments; 2. How to design experiments to detect the presence of a scalar field constituting dark energy, especially by searching for new force between objects separated by distances below submillimeter scale.

One of the theoretical scenarios is the chameleon field proposed by Khoury and Weltman, in which the mass of the field depends on the local energy density (*6-8*). The effective mass corresponds to the inverse length scale of the interaction. In high-density environments, such as on Earth, the mass is large and the length of the interaction is short, which makes it difficult to test a fifth force mediated by the chameleon field. In low-density environments, such as the cosmological distances, the mass is small and then the chameleon field would mediate a long-range interaction that explains the accelerating expansion of the universe. Since most laboratory experiments are carried out in regions of high matter energy density, the forces arising from the chameleon field are strongly suppressed. Recent precision experiments in ultrahigh-vacuum situation have shown more constraints on chameleon scalar fields (*9, 10*). But how to test if the chameleon fields actually exist, and whether the chameleon theory can naturally and quantitatively explain the density of dark energy, remain as the most pressing open questions in physics.

To test the chameleon force in a laboratory, at least two objects are needed. One is used to perturb the chameleon field, called source object; the other one to test the field, called test object. The severe challenge is how to decrease the distance between the two objects in order to avoid the strong screening effect of the chameleon field. One of the essential characteristics of chameleon screening effect is that the chameleon force is only strong in a thin shell near the surface of a source object. Compared to the force due to gravity the chameleon force on a test mass would be very large near the surface but would be strongly suppressed away from the surface. Previous searches for violations of Newtonian gravity in the submillimeter regime employed at least one macroscopic object as source or test mass. The object often has to be mounted on the other solid matter (*9, 10*). This contact may change the chameleon interaction, just like how electrical contact can change electrostatic screening effects. The thin-shell effect of the chameleon field is analogous to a thin layer of charge residing on a conductor surface (*4, 11*).

In this study, we used two atom clouds of rubidium-87 as our test mass and source mass in an ultrahigh vacuum chamber (Fig.1).The source atom cloud is a part of background rubidium-87 atoms covered by probe laser beams, *i.e.*, the configuration of one-dimensional (1-D) optical molasses (*12-15*). Besides the traditional use of the optical molasses technique to decrease the energy of the atom cloud, we also used the configuration to increase its energy as long as the laser frequency is tuned above the atomic resonance (or the blue-detuning). Therefore, the energy density of the source mass can be adjusted by probe laser. The test atom cloud is formed by a standard laser cooling technique known as three-dimensional (3-D) optical molasses (*12-15*). We released the test atom cloud downward to the source atom cloud and measured the fall acceleration of the test atom cloud with the time of flight (TOF) method (*14-18*). This properly designed scheme can not only suppress screening effect due to the low density of ultrahigh-vacuum chamber, but also allow the distance between the two masses to approach zero because both masses are atom clouds. Since the test mass can pass through the source mass in space, the thin-shell effect may be uncovered and the chameleon force may be detected.

The test mass formed by the 3-D optical molasses was suspended in the middle of the vacuum cell. The energy density of the source mass is adjusted by tuning the frequency of the probe light in the 1-D molasses configuration. When an atomic system in an equilibrium state is illuminated by laser beams, an adjustable energy density source for exciting the chameleon field is generated. The source mass is almost isolated from the walls of the vacuum cell except for the two ends of the source, but the cross-section areas are very small compared to its flank area. Both the test mass and source mass suspended inside of the vacuum chamber were almost isolated from the other dense matter, such as the solid walls of the vacuum cell. Such isolation may be another key element in testing the chameleon field, just like an insulating measure in an electrostatic experiment. We observed the small fall acceleration of the falling atom cloud, comparing with the free-fall acceleration due to gravity when the probe laser frequency was tuned below the atomic resonance (or the red-detuning situation). In the situation of the probe light with relatively larger blue-detuning frequency, the measured acceleration exceeded the gravitational acceleration. We attribute the detuning-dependent slowing and quickening of the test atom clouds to the chameleon fields around the cylindrical matter energy source controlled by the pair of laser beams.

The dynamics of the chameleon dark energy field $\phi$ having the dimension of energy are governed by an effective potential density $V_{\text{eff}}(\phi)$ (*6*), *i. e.*

$$\hbar^2 c^2 \nabla^2 \phi = \frac{d}{d\phi} V_{\text{eff}}(\phi) \qquad (1)$$

where $\hbar$ is the recued Planck constant, $c$ is the speed of light. The effective potential density $V_{\text{eff}}(\phi) = V(\phi) + V_{\text{int}}$ which is a sum of a self-interaction $V(\phi)$ and an interaction $V_{\text{int}}$ with ordinary matter. We will use commas to denote derivatives, $e.g., V_{,\phi} = dV(\phi)/d\phi$. The interaction potential density is an explicit function of local matter density $\rho$ (6),

$$V_{\text{int}} = \rho \hbar^3 c^5 e^{\beta \phi / (M_{\text{Pl}} c^2)} \qquad (2)$$

where the reduced Planck mass $M_{\text{Pl}} = (\hbar c / 8\pi G)^{1/2} \approx 2.4 \times 10^{18}$ GeV/c$^2$ with the gravitational constant $G$. $\beta$ is a dimensionless coupling of the scalar field to matter. The coupling can also be characterized by another parameter with the dimension of mass, that is $M \equiv M_{\text{Pl}}/\beta$. The coupling parameter is essentially not constrained till now (9, 10). One of the reasons is that the properties of a chameleon field depend strongly on the shape of the bare potential $V(\phi)$. The runaway form, such as

$$V(\phi) = \Lambda^4 e^{\Lambda/\phi} \qquad (3)$$

with $\Lambda \approx 2.4$ meV, and the quartic self-interaction form are frequently used (6-11). The quartic self-interaction form is (7)

$$V(\phi) = \frac{\xi}{4!} \phi^4 \qquad (4)$$

where $\xi$ is a dimensionless parameter. We will set the "natural" value that $\xi = 1$. Although the concrete form of self-interaction is unknown, using reasonable physical analysis one can approximately derive acceleration expressions of test objects due to the chameleon fields (11). For example, in the case of the chameleon field around a cylindrical source, the maximum of the acceleration occurring at the surface of the cylindrical source is given by (19)

$$a_{\max} \approx -4\pi G \rho_c \beta^2 \sigma_p \qquad (5)$$

where $\sigma_p$ is the radius of the cylindrical source, $\rho_c$ is the mass density of the source. The chameleon field pulls a test mass toward the center of the source denoted by the minus sign, and the attractive force reaches its maximum at the surface of the cylindrical source, which coincides with the thin-shell effect from the chameleon field.

Our setup can be easily replicated in laboratories on Earth (18). The test atom cloud including cold atomic number $\sim 4 \times 10^8$ and with temperature $\sim 17$ μK was prepared in the ultrahigh vacuum of pressure $\sim 2 \times 10^{-7}$ Pa. The initial size of the atom cloud $\sigma_0 = 0.8$ mm was determined by fluorescence imaging with a charge-couple-device camera (CCD). The mass density of the test atom cloud was estimated to be $\sim 2.7 \times 10^{-8}$ kg/m$^3$ which is much smaller compared to the density of the atmosphere. We used a pair of cylindrical laser beams located 7.5 mm below the center of the test atom cloud to excite the background atoms to form a source object (the distance was calibrated by CCD photographs of the 3-D optical molasses location and probe light location). The density of background atoms in the vacuum cell was adjusted to the desired status by running the current through a dispenser (not shown in Fig.1). The pair of counter-propagating laser beams was also used to excite the falling test atom cloud to get TOF signals. Therefore the laser beams were called probe light. The TOF signal of the ballistically expanded cloud was obtained by collecting the fluorescence of the falling atomic cloud excited by the probe light. The probe beams with Gaussian radius of $\sigma_p = 1.3$ mm had a fixed power $P_0 \simeq 203$ µW per beam.

When the cooling beams of the test atom cloud were switched off, the test atom cloud began to fall due to gravity. At the same time, the probe laser light was switched on. The probe beams were endowed with twofold functions mentioned above: one is for generating the source mass; the other for exciting the falling test mass in getting TOF signals. We detected the fall acceleration of the test atoms by monitoring their fluorescence with a photodiode (PD). By recording the fluorescence excited by the probe light, the fall acceleration of the atomic cloud was obtained.

Varying the frequency of the probe beams while fixing their power, we obtained the detuning-dependence of the fall accelerations of the test atom clouds. The detuning $\Delta = \omega_p - \omega_0$ is the probe light frequency $\omega_p$ from the atomic resonance frequency $\omega_0$. Figure 2 shows the fall accelerations versus the detuning of the probe beams, where the circles are the experimental data. In the red-detuning regime, the fall accelerations were lowered relatively to the free-fall acceleration $g$ (9.794 m/s² at Shanghai, China), which could be attributed to the radiation pressure of the fluorescence of the background atoms (RPFB) as analyzed in (18). When the laser frequency was tuned above the atomic resonance and was gradually increased, the measured fall acceleration began to increase and even exceeded the free-fall acceleration. This can no longer be explained by the RPFB. Because the probe beams located below the test atom cloud, the radiation pressure was only able to lower rather than to heighten the fall accelerations. The dotted curve shown in Fig. 2 is obtained by considering only the contribution of the RPFB besides gravity, i.e., $a'_{tot} = g - a_f$ (18), where $a_f \propto (4\Delta^2 + \Gamma^2)^{-1}$ is the fractional part of the fall acceleration due to the RPFB with $\Gamma$ being the spectral width of the atom. The dotted curve is symmetric about the axis of $\Delta = 0$, which is inconsistent with the unsymmetrical experimental data (circles in Fig. 2).

To ensure little variation in the experimental conditions we recorded data one laser frequency point after another. Thus, the unsymmetrical experimental data about the detuning values, especially the acceleration enhancement firmly shows a new physics. We will analyze the experimental data by the chameleon forces in the latter part of this paper (20).

We have pointed out that the energy densities of source atom clouds were adjusted by probe laser beams. When the laser beams irradiate the background atomic system in the ultrahigh vacuum chamber, there exists a net energy exchange between the atomic system and light field due to spontaneous emission. When the laser frequency is tuned above the atomic resonance, the atomic system obtains positive energy from the photon field, and vice versa (12). Then the mass density of the source atoms is estimated as

$$\rho_c = \rho_{bg} + \frac{\delta E}{c^2} \qquad (6)$$

where $\rho_{bg}$ is the mass density of the background atom gas, $\delta E$ is the increased energy density of the atom system and $\delta E \propto \Delta \cdot (1 + 4\Delta^2/\Gamma^2 + I/I_{sat})^{-2}$ (Eq. S5) with probe light intensity $I$ per beam and saturation intensity $I_{sat}$. The mass density of the source increases compared to that of the background atom gas when the driving laser frequency is tuned above the atomic resonance, i.e. $\Delta > 0$, and vice versa. Physically, the region of source mass will pull the test mass in the case of the blue-detuning of $\rho_c > \rho_{bg}$, while the region of the source mass will push the test mass in the case of the red-detuning of $\rho_c < \rho_{bg}$. Mathematically, if the probe beams have not irradiated the background atoms, the atomic system is homogeneous and then $\nabla^2 \phi = 0$ or $V_{,\phi}(\phi) = -V_{int,\phi} \approx -\rho_{bg} \hbar^3 c^3 \beta / M_{Pl}$. At the higher density of $\rho_c > \rho_{bg}$, the effective chameleon potential *beneath* the source surface is well approximated by $V_{eff}(\phi) \approx V_{int}$ (*19*) and $V_{eff,\phi}(\phi) \approx V_{int,\phi} \approx \rho_c \hbar^3 c^3 \beta / M_{Pl}$; At the lower density of $\rho_c < \rho_{bg}$, the potential *on* the source surface is well approximated by $V_{eff}(\phi) \approx V_{int}$ and $V_{eff,\phi}(\phi) \approx V_{int,\phi} \approx \rho_{bg} \hbar^3 c^3 \beta / M_{Pl}$. Although the energy transferred to the source mass from the laser beams is rather small, *i.e.*, $\rho_c \approx \rho_{bg}$, the chameleon forces will change their directions when the detuning of the driving laser varies from negative to positive. This direction reversal ensures us to detect the new force. Then Eq. (5) can be generalized approximately as (*19, 20*)

$$a_{max} \approx \mp 4\pi G \cdot \max(\rho_c, \rho_{bg}) \cdot \beta^2 \sigma_p \qquad (7)$$

'−' and '+' correspond to $\rho_c > \rho_{bg}$ and $\rho_c < \rho_{bg}$, respectively; $\max(\rho_c, \rho_{bg})$ denotes the biggest of $\rho_c$ and $\rho_{bg}$. We use the sign function and the approximation $\rho_c \approx \rho_{bg}$ to rewrite Eq. (7) as $a_{max} \approx -\text{sgn}(\rho_c - \rho_{bg}) 4\pi G \rho_c \beta^2 \sigma_p$. In order to fit the experimental data quantitatively, we use roughly half of $a_{max}$ as an average value of the chameleon acceleration $a_\phi$, and attempt to use a smoothly analytic approximation of the sign function $(2/\pi) \cdot \tan^{-1}(\Delta/\Gamma)$, *i. e.*,

$$a_\phi = -\frac{2}{\pi} \tan^{-1}\left(\frac{\Delta}{\Gamma}\right) \cdot \frac{1}{2} \cdot 4\pi G \rho_c \beta^2 \sigma_p \qquad (8)$$

The solid curve in Fig. 2 is calculated by the total acceleration $a_{tot} = g - a_f - a_\phi$, in which the mass density of rubidium -87 is $\rho_c \approx 7 \times 10^{-12} \text{ kg/m}^3$ corresponding to the vacuum pressure $2 \times 10^{-7}$ Pa. We estimated relatively precise mass density of the vacuum chamber via measuring the life time of cold atom cloud in the magnetic trap. As a result, the most important parameter $\beta$ is fitted as

$$\beta \approx 4.5 \times 10^{11} \qquad (9)$$

Then the coupling parameter with the dimension of mass is

$$M \equiv \frac{M_{Pl}}{\beta} \approx 5.3 \times 10^6 \text{ GeV}/c^2 \qquad (10)$$

Since the use of the half of $a_{max}$ as an average value of $a_\phi$ overestimates the acceleration due to the chameleon field, the more accurate value of $\beta$ will be larger than $4.5 \times 10^{11}$. This is a strong interaction. It has been shown that chameleon mechanism allows scalar fields to couple to matter much more strongly than gravity does, and still satisfies the current experimental and observational constraints (*8*). We will go much further and show, unlike runaway potential for $\phi$ where the strength of the self-interaction is equivalent to the density of dark energy (Eq. (3)), that the interaction $V_{int}$ potential is also possible to connect to dark energy in a quartic self-interaction for $\phi$. The length scale of the chameleon force with the quartic self-interaction is (*7*)

$$l_\phi \equiv \frac{\hbar}{m_{eff} c} = \frac{\sqrt{2}}{\xi^{1/6}} \left(\frac{M_{Pl}}{6\beta\rho}\right)^{1/3} \qquad (11)$$

where $m_{eff}$ is the mass of fluctuation around the minimum of the effective potential. Assuming $\xi = 1$, we find that the range of chameleon-mediated interactions in the atmosphere is about 0.2 $\mu$m ($\rho_{atm} \approx 1$ kg/m$^3$ (*21*)), in the solar system is about 2 m ($\rho_G \approx 10^{-21}$ kg/m$^3$ (*21*)) and on the cosmological scales today is about 120 m (using the current matter density $\rho_0 \approx 2.7 \times 10^{-27}$ kg/m$^3$ of the universe including dark matter (*2*)). These length sales are very short even for the very low density. 120 m is an average length scale for the average matter density in a huge region of the universe. At cosmological distances there exist lots of empty regions where the matter densities approach zero and the chameleon field would mediate a long-range force. The empty regions would push the dense matter away that is similar to the red-detuning situation in our experiment mentioned above. The chameleon force originates from the spatial dependence of the field, *i.e.*, the gradient. We cannot use this very inhomogeneity to explain the accelerating expansion of the universe due to the expansion being global and homogeneous. The accelerating expansion of the universe is attributed to the pressure density which is related to the potential density itself rather than its first and second derivatives with respect to the field. Since the coupling between the matter and the chameleon field is dependent not only on the matter density but also on the chameleon field, it is possible that the interaction potential with ordinary matter could be regarded as a part of the total potential of the chameleon field. The contribution of the interaction potential to the energy density and the pressure density in the equation of state of the chameleon field could be taken into account. Then considering a homogeneous chameleon field in space, its energy density and the pressure density can be described, respectively, as (*22-31*)

$$\rho_\phi = \frac{1}{2}\hbar^2 \dot{\phi}^2 + V_{eff}(\phi) \tag{12}$$

$$P_\phi = \frac{1}{2}\hbar^2 \dot{\phi}^2 - V_{eff}(\phi) \tag{13}$$

where overdots denote derivatives with respect to coordinate time $\dot{\phi} = \partial \phi / \partial t$. The equation of state $w = P/\rho$ is given by

$$w_\phi = \frac{\hbar^2 \dot{\phi}^2 / 2 - V_{eff}(\phi)}{\hbar^2 \dot{\phi}^2 / 2 + V_{eff}(\phi)} \tag{14}$$

It should be emphasized that in these equations (12) - (14) we have used $V_{eff}(\phi)$ including interaction with matter instead of the bare potential $V(\phi)$. When the total potential energy $V_{eff}(\phi)$ dominates over the kinetic energy $\hbar^2 \dot{\phi}^2 / 2$, the accelerating expansion of the universe occurs. For the case of $\hbar^2 \dot{\phi}^2 / 2 \ll V_{eff}(\phi)$, we have $w_\phi \approx -1$ and the energy density $\rho_\phi \approx V_{eff}(\phi)$. Since the field value corresponding to the minimum of the effective potential is $\phi_{min} = \left[-6\hbar^3 c^3 \beta \rho / (\xi M_{Pl})\right]^{1/3}$ in the case of $\beta > 0$ (*7*), the minimum value of the energy density is dominated by $V_{int}$ and estimated as $V_{eff}(\phi_{min}) \approx \rho \hbar^3 c^5$ (*32*). This means that the chameleon's dark energy density equals to the matter energy density including dark matter. If we define an energy scale of $\Lambda_\phi$ corresponding to the scalar field by the minimum of the effective potential, *i.e.*,

$$\Lambda_\phi^4 \equiv V_{eff}(\phi_{min}) \tag{15}$$

the value of $\Lambda_\phi$ is estimated to be about 1.8 meV for matter density of $\rho_0 \approx 2.7\times 10^{-27}$ kg/m$^3$. The definition including the interaction potential differs from that in literature (*22-31*). For example, the bare self-interaction potential's value at this minimum is always regarded as the role of cosmological constant (or, equivalently, dark energy) (*31*). Consequently, the energy scale corresponding to the cosmological constant has to be introduced at the beginning indicated by Eq. (3). The current dark energy density is about twice as much as the matter energy density including dark matter (*2*). This may indicate that there are two chameleon potential minima in our universe. Or, it is possible that the chameleon field couples matter with both $\beta > 0$ and $\beta < 0$ in some way (Fig. 3, A and B).

In conclusion, we have detected a fifth force between the two atom clouds using the cold atom technique. When the energy density of the source atom cloud is larger than that of the background, the test mass is pulled to the center of the source mass, and vice versa. Based on the chameleon theory, the coupling constant of $4.5\times 10^{11}$ is obtained by fitting the experimental data. Employing the quartic self-interaction, the strong coupling to matter explains why the most sensitive previous searches cannot detect the presence of the chameleon forces. Although the forces are considerably larger than Newtonian gravity in short distances, the interaction ranges are short enough to satisfy all experimental bounds on deviations from general relativity. The strength of the chameleon force is strongly relying on the first derivative of the potential density with respect to the chameleon field and spatial gradient. Thus, the larger the coupling constant $\beta$ is, the stronger the chameleon force. The interaction ranges are related to the second derivative of the potential density with respect to the chameleon field. The larger the coupling constant $\beta$ is, the shorter the interaction ranges. When the contribution of the interaction potential to the energy density and the pressure density of the scalar field is taken into account, a dark energy scale of 1.8 meV is estimated. The estimated value corresponding to the cosmological constant is insensitive to the value of the coupling constant.

constant. *Phys. Rev. D* **93**, 124003 (2016).
32. P. Brax, C. Burrage, Atomic precision tests and light scalar couplings. *Phys. Rev. D* **83**, 035020 (2011).

**Acknowledgments:**

We acknowledge discussions with J. N. Zhang, C. Liu, Y. Z. Wang, P. F. Zhang, X.P. Xu, B. L. Zhang, X. Y. Yu and J. H. Wen. This work was supported by the National Natural Science Foundation of China through Grant Nos. 10474105.

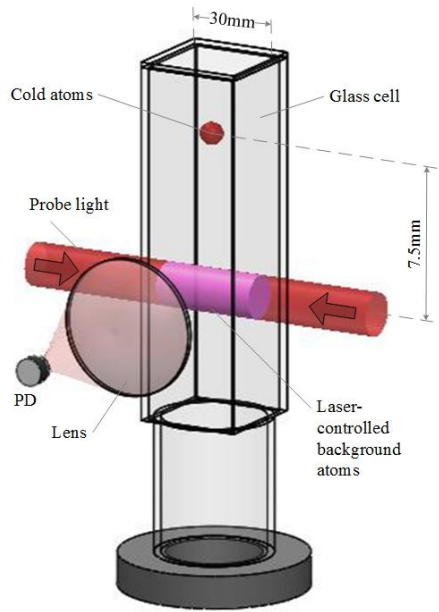

**Fig1. Schematic of our experimental setup**. The density of background rubidium-87 atoms in the vacuum cell (30 mm × 30 mm × 70 mm) was adjusted to a desired pressure by running the electric current through a dispenser (not shown). The source mass was a part of the background atoms which was excited by the pair of cylindrical probe laser beams, and its energy densities were adjusted by tuning the laser frequencies. The test mass was formed by a standard laser cooling technique, which was located 7.5 mm above the center of the probe light. The fall accelerations of the test mass were measured by the TOF method. The TOF signals were recorded with a photodiode (PD) by collecting the fluorescence of the falling test atoms excited by the probe light.

The lens ( $f = 80$ mm, $\Phi = 30$ mm ) was used to collect a lager solid angle range of the fluorescence.

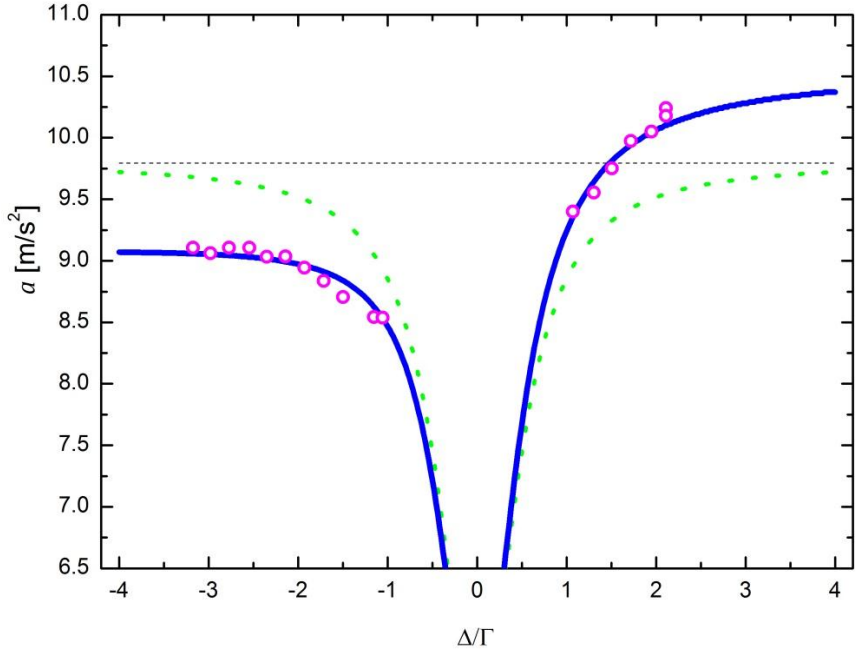

**Fig.2. The fall acceleration versus the detuning $\Delta/\Gamma$.** The circles are experimental data points obtained by fitting the TOF signals using Eq. (S6). The solid curve is calculated by $a_{tot} = g - a_f - a_\phi$ to fit the experimental data, where $a_f$ is calculated with the fitting parameter of the fluorescence intensity $I_f = 0.15$ $\mu$W/cm$^2$ resulted from the background atoms, and $a_\phi$ is calculated by Eq. (8) with the fitting parameter $\beta \approx 4.5 \times 10^{11}$ and the measured density $\rho_c \approx 7 \times 10^{-12}$ kg/m$^3$. The dotted curve is obtained by $a'_{tot} = g - a_f$ without considering $a_\phi$, with the same calculated parameters mentioned above. The saturation intensity is $I_{sat} = 3.576 \text{ mW/cm}^2$.

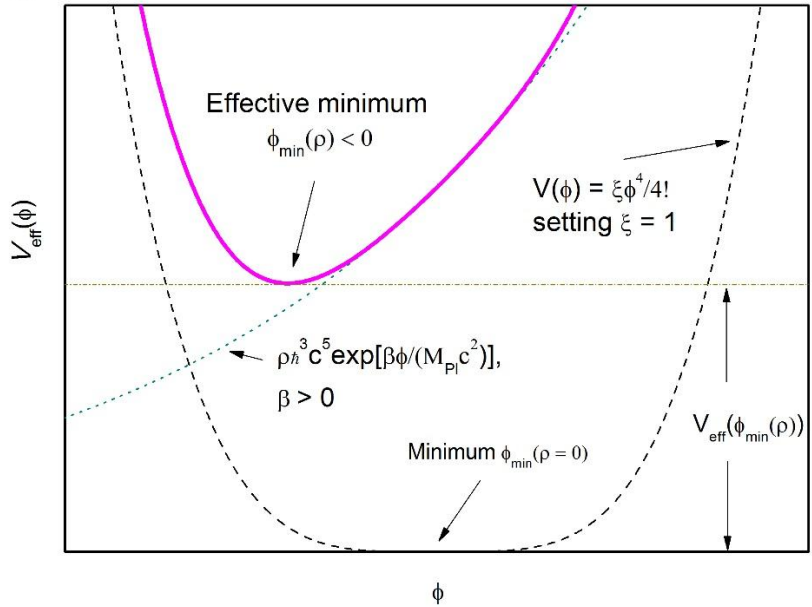

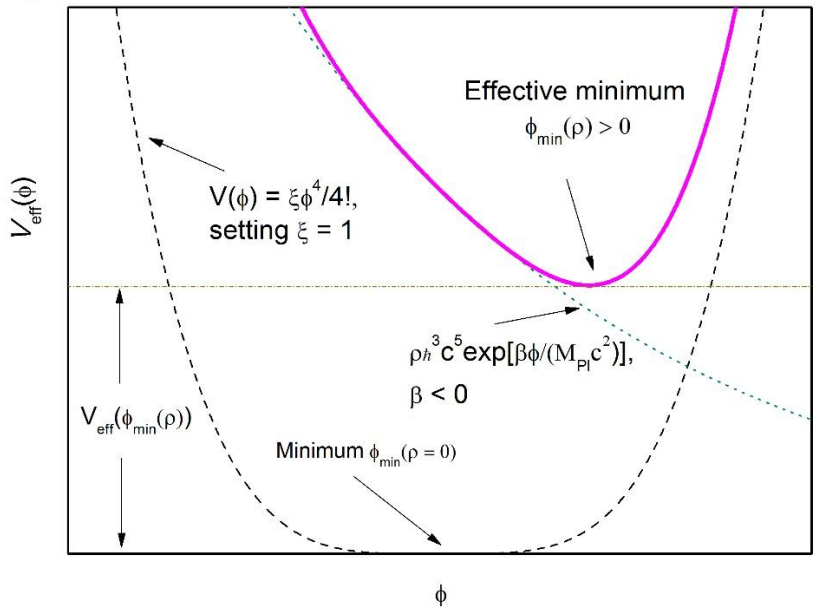

**Fig. 3. Chameleon effective potential density versus chameleon field $\phi$.** For a positive coupling constant (A), the effective potential density is minimized at a negative value of $\phi$ and the minimum of the potential density in the homogeneously cosmological scale defines a chameleon's dark energy density. For a negative coupling constant (B), the effective potential is minimized at a positive value of $\phi$ and the corresponding chameleon's dark energy density is the same as above.

**Supplementary Materials:**

Supplementary Text

Figs. S1 to S2

Equations S1 to S9

**Supplementary Materials:**

**Supplementary Text**

The chameleon mechanism

The equilibrium value of $\phi$ in the presence of matter density $\rho$ is the minimum $\phi_{min}$ of the effective potential $V_{eff}(\phi)$. Small fluctuation around this minimum can be described by the mass term arising from a harmonic expansion of the potential $V_{eff}(\phi)$. The mass of the chameleon field corresponds to the second derivative of the effective potential with respect to the chameleon field, $m_{eff}^2 c^4 = \partial^2 V_{eff}(\phi)/\partial \phi^2$. Based on the harmonic approximation and the thin shell effect of the chameleon field one can give approximated solutions to the equation of motion (Eq. (1)) without the concrete expression of the bare potential $V(\phi)$. For example, the chameleon field around a cylindrical source is discussed in (*19*). The chameleon field pulls a test object at a radius $r$ from the center of the cylindrical source with acceleration (Eq. (A. 29) in (*19*))

$$a_\phi = -\frac{\beta}{M_{Pl}}\frac{\partial \phi}{\partial r} \simeq -\frac{\hbar c \rho_c \beta^2 \sigma_p^2}{2M_{Pl}^2 r}, \quad r > \sigma_p \tag{S1}$$

The maximum value of the acceleration occurs at the surface of the cylindrical source, $a_{max} = -\hbar c \rho_c \beta^2 \sigma_p / 2M_{Pl}^2 \equiv -4\pi G \rho_c \beta^2 \sigma_p$.

Setup

Owing to the thin-shell effect, the sharp variation of the chameleon field in space coordinates only occurs near the surface of the source mass, which results in a large chameleon force near the surface but a great suppression in the rest of the region. Therefore, we adopt the strategy of putting sensors into the region of the thin shell instead of just only improving the diagnostic sensitivity while unable to put the sensors into the region.

We used two atom clouds to detect the chameleon force. Our setup is a standard cold atom experiment system which has been described in (*18*). The test mass is the cold atom cloud prepared in a conventional laser cooling technique. The source mass is a part of the background rubidium-87 atoms whose energy density was adjusted by the probe light. We only give the energy density of the source mass here. When the laser beams irradiate the background atomic system in the ultrahigh vacuum chamber, there exists a net energy transfer between the atomic system and light field due to spontaneous emission. This configuration is substantially a 1-D molasses (*12*). In this configuration, the total friction force exerted on an atom in the two energy-level transition is

$$F = -\alpha v_z \tag{S2}$$

where $v_z$ is the absolute value of the atom velocity projection onto the laser propagating direction, and

$$\alpha = 4\hbar k^2 \cdot \frac{I}{I_{sat}} \cdot \frac{\Delta/\Gamma}{\left(1 + 4\Delta^2/\Gamma^2 + I/I_{sat}\right)^2} \tag{S3}$$

with $k$ being wave vector of the probe light, $I = P_0 / (\pi \sigma_p^2)$ is the intensity of the probe light per beam. The energy changed per unit time is

$$W \sim -F v_z \tag{S4}$$

The mean value of the density of the transferred energy is

$$\delta E \sim t_d \cdot \iiint W f(v_x, v_y, v_z) dv_x dv_y dv_z = \alpha t_d n \frac{k_B T}{m} \tag{S5}$$

where $f(v_x, v_y, v_z)$ is the Maxwell distribution function, $m$ is the atomic mass, $k_B$ is the Boltzman constant, $n = \rho_{bg}/m$ is the atomic number density of the background atom gas, $t_d$ is the interaction time between the probe laser light and the atomic system, $T$ is the temperature of the background atom gas. Then the total mass density of the source is $\rho_c = \rho_{bg} + \delta E/c^2$, which depends on the detuning $\Delta$ indicated by Eqs. (S5) and (S3). The term $\delta E/c^2$ is much smaller than $\rho_{bg}$ ($\rho_c \approx \rho_{bg}$).

Although the absolute value of the source density changes slightly, the most important function of the detuning dependence of $\rho_c$ is that, the chameleon force will switch its direction when the sign of the detuning is shifted. When $\delta E > 0$, the chameleon field pulls a test object toward the center of the cylindrical source, and the maximum value of the acceleration occurring at the surface of the source is $a_{max} = -\hbar c \rho_c \beta^2 \sigma_p / 2M_{Pl}^2$. When $\delta E < 0$, the chameleon field pushes a test object away from the center of the cylindrical source and the maximum value of the acceleration occurring at the surface of the source is estimated as $a_{max} = +\hbar c \rho_{bg} \beta^2 \sigma_p / 2M_{Pl}^2$. When $\delta E = 0$, the chameleon force equals to zero. In other words, the direction of the chameleon force always points to the dense region and $a_{max} = -\text{sgn}(\delta E) \max(\rho_c, \rho_{bg}) \hbar c \beta^2 \sigma_p / 2M_{Pl}^2$. In our case that $\rho_c \approx \rho_{bg}$, for simplicity we rewrite this expression above as follows $a_{max} \approx -\text{sgn}(\rho_c - \rho_{bg}) \hbar c \rho_c \beta^2 \sigma_p / 2M_{Pl}^2 \equiv -\text{sgn}(\rho_c - \rho_{bg}) 4\pi G \rho_c \beta^2 \sigma_p$. Figure S1 shows the sketch of the thin shell of the chameleon field in both cases.

As a further supplement, it should be emphasized that we cannot use Eq. (S1) to describe the acceleration in the case of $\rho_c < \rho_{bg}$, or we will conclude a wrong result that the region still pulls the test object toward its center just like in the case of $\rho_c > \rho_{bg}$. Using the same expression in both situations neither satisfies the chameleon theory nor agrees with our experiments. Also, we cannot take it for granted that using $\rho_c - \rho_{bg}$ to replace $\rho_c$ in Eq. (S1) or in Eq. (7) while solving the equation of motion for the chameleon. By contrast, we temporarily assume that $a''_{max} = -\hbar c(\rho_c - \rho_{bg})\beta'^2 \sigma_p / 2M_{Pl}^2 \equiv -4\pi G(\rho_c - \rho_{bg})\beta'^2 \sigma_p$ and use $a''_{tot} = g - a_f - a''_\phi$ to fit the experimental dada, where $a''_\phi$ is half of $a''_{max}$ and other parameters are same as in Fig. 2. Figure S2 shows the fall accelerations versus the detuning of the probe beams with the false assumption of $a''_\phi = -4\pi G(\rho_c - \rho_{bg})\beta'^2 \sigma_p / 2$ to fit the experimental dada (circles). The solid curve in Fig. S2 calculated by $a''_{tot} = g - a_f - a''_\phi$ cannot be adjusted to agree with the experimental dada.

Time of flight method

It is very important to analyze the time-of-flight (TOF) method in the specific case of short distances of the probe beams from the initial position of the atom cloud. This method can be used not only to measure the temperature of the atom cloud, but also to estimate the fall acceleration. The TOF signal of the ballistically expanded cloud was obtained by collecting the fluorescence of the falling atomic cloud excited by the probe light. According to Brzozowski *et. al. (17)*, the TOF signal at time $t$ is given as follows

$$S(t) \propto \frac{P_0}{\sigma_0^2 + \sigma_v^2 t^2 + \sigma_p^2} \exp\left\{-\left(\frac{a(t_0^2 - t^2)}{2\sqrt{2}\sqrt{\sigma_0^2 + \sigma_v^2 t^2 + \sigma_p^2}}\right)^2\right\} \tag{S6}$$

Where $\sigma_0$ is the initial Gaussian radius of the expanded atom cloud; $\sigma_v$ is the Gaussian radius of the velocity distribution of the atom cloud, which associated its temperature by the formula of $T_{test} = m\sigma_v^2 / k_B$; $a$ is the fall acceleration of the atomic cloud; $t_0 = \sqrt{2d/a}$ is the arrival time of the center of the test atom cloud without initial vertical velocity and $d$ is the distance between the initial position of the test atom cloud and the location of the probe light. In our experiment, the interaction time $t_d$ between the probe laser light and the atomic system mentioned above is designed to be equal to $t_0$. A typographical error in Eq. (12) of the original literature *(17)* is that $\sigma_p$ in the exponent was omitted, which has been added in Eq. (S6) here.

In our experiment, the probe beams with Gaussian radius of $\sigma_p = 1.3$ mm had a power $P_0 \approx 203$ $\mu$W per beam. Varying the frequency of the probe beams but fixing their power, we collected the TOF signals 10 times for each laser frequency point and arithmetically averaged them. We fitted the signal curves by using Eq. (S6) and obtained the fall accelerations which are detuning dependent. The large range of the detuning was not achieved due to the limitation of our acousto-optical modulator (AOM). In the very near resonance regime, the TOF signals are too noisy to be fitted from the experimental TOF curves in obtaining the accelerations.

Radiation pressure of the fluorescence of the background

The fluorescence emitted from the background atoms can influence the TOF signal, which was discussed in (*18*). The irradiation pressure to the test atoms depends on the detuning. The detuning-dependent acceleration $a_f$ is upward whether the detuning is negative or positive (*18*), *i.e.*,

$$a_f = \frac{\hbar k \Gamma^3}{2m I_{sat}} \cdot \frac{I_f}{4\Delta^2 + \Gamma^2} \tag{S7}$$

The intensity of the fluorescence $I_f$ is much smaller than the saturation intensity $I_{sat}$. Then $I_f/I_{sat}$ shown in the denominator of Eq. (7) in (*18*) is safely neglected here (Eq. S7). It is possible to choose a fitting parameter of $I_f$ to approach the experimental data in the red-detuning case. But it is totally impossible to do that in the blue-detuning case because the quickening accelerations occur while $a_f$ is upward. Besides, it is not possible to find a compromise parameter to satisfy both cases due to $a_f$ being symmetric about $\Delta = 0$. By using the chameleon mechanism, all the fitting parameters are almost determined uniquely, that is, there is no wide range to choose the fitting parameters from due to the simultaneous equations. Each experimental datum stands for an equation. Mathematically, there are many redundant equations to constrain the unknown quantities. In this context, $d$ (the distance between the initial position of the test atom cloud and the location of the probe light) can also be obtained as a fitting parameter.

If chameleon forces did not exist, the measured fall accelerations would have the same value for blue-detuning and red-detuning as long as the absolute values of the detuning were equal. This is not the experimental case. If one just had the data in the blue detuning situations only, one could not have the confidence to attribute the exceeded accelerations to a new type force. One tends to think that the quickening acceleration is a false appearance and to regard it as some systematic errors, for example, the uncertainties of the distance between optical molasses and probe light, the falling time of the test atom cloud, and so on. The data in the red detuning case can be used to remove the influence due to the RPFB.

One may argue that $a_f$ is slightly larger for blue-detuning compared to that for red-detuning due to $\omega_p = ck$. But the slightly large wave vector only means that $a_f$ is larger rather than smaller in the case of the blue-detuning. We also cannot attribute the extra acceleration $a_\phi$ to the optical dipole force formed by the probe laser beams. This gradient force not only is too weak but also gives a reversed effect compared to the experimental results. That is, for the red-detuning the force drives the atoms to positions where the light intensity has a maximum, whereas for the blue-detuning the force pushes the atoms away from the intensity maximum.

The length scale of the chameleon force

We adopt the bare potential in the quartic self-interaction form. The equilibrium value of $\phi$ is the minimum $\phi_{min}$ of $V_{eff}(\phi)$. Using the linear approximation, one has (*7*)

$$\phi_{min} = \left( \frac{-6\hbar^3 c^3 \beta \rho}{\xi M_{Pl}} \right)^{1/3} \tag{S8}$$

$$m_{eff} = \frac{-\xi^{1/2}}{\sqrt{2}c^2} \phi_{min} \tag{S9}$$

Where minus sign results from the choice of $\beta>0$. One could equivalently consider the case $\beta<0$. For the global universe, the two cases $\beta>0$ and $\beta<0$ may exist together in some way which cannot be identified in this study. When we choose a natural value $\xi=1$, we find that the interaction range is very short even in the commonly high vacuum situation due to $\beta\sim10^{11}$ rather than $\beta\sim1$. This makes it difficult to detect the chameleon force even for a sensitive experimental design. The length scale of the chameleon force inside our vacuum cell ($\rho_{bg}\approx7\times10^{-12}\,\text{kg/m}^3$) is about 0.9 mm, which is much smaller than the cell width of 30 mm. This means that the chameleon field adapts to the change of density between the cell walls and the vacuum in a very short distance. The field at the center of the chamber can adapt to the vacuum value of the chamber, and the influence of the cell walls on the field in the middle of the chamber can be safely neglected.

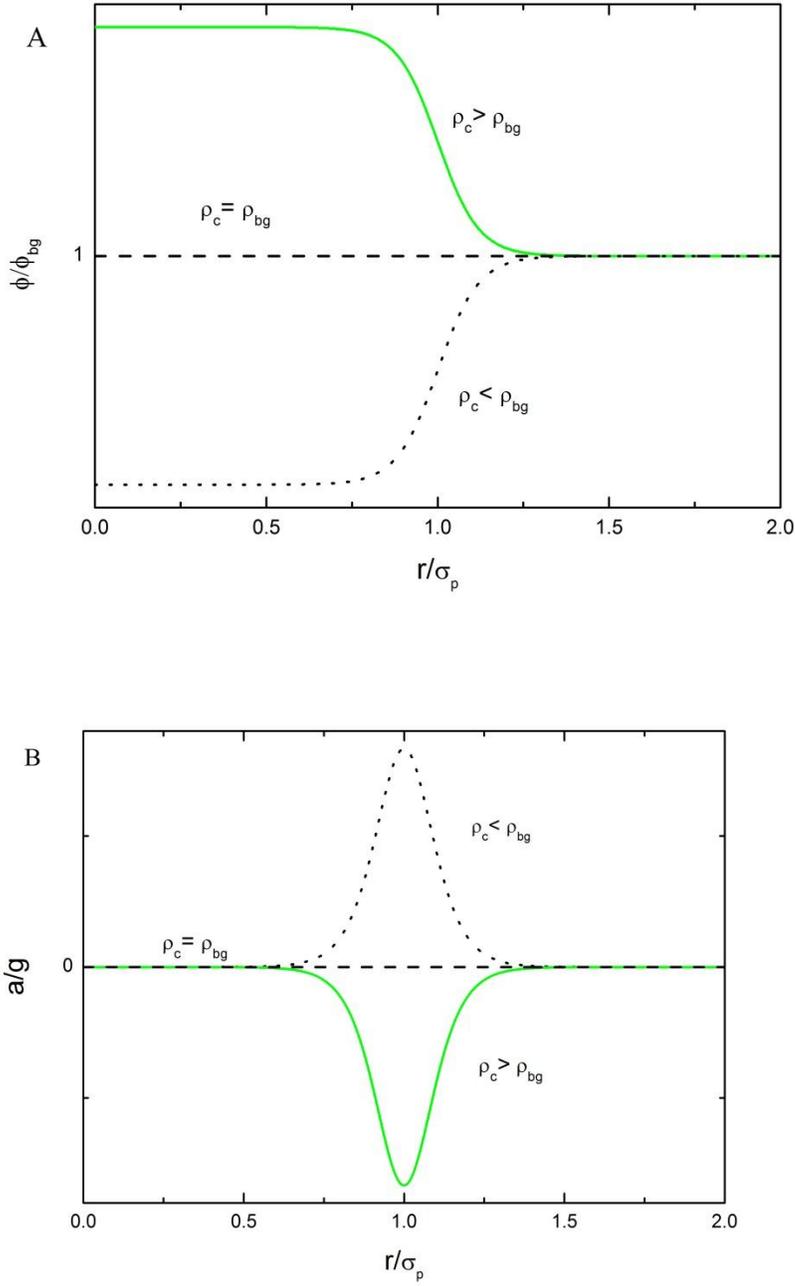

**Fig. S1. Sketch of thin shell of chameleon field.** Profile of chameleon field varies sharply near the surface of the source mass (A), which results in a large chameleon acceleration near the surface but a great suppression in the rest of the region (B). The chameleon acceleration will reverse its sign when the difference of $\rho_c - \rho_{bg}$ varies from positive to negative. This direction reversal effectively uncovers the presence of the chameleon field and ensures us to detect it.

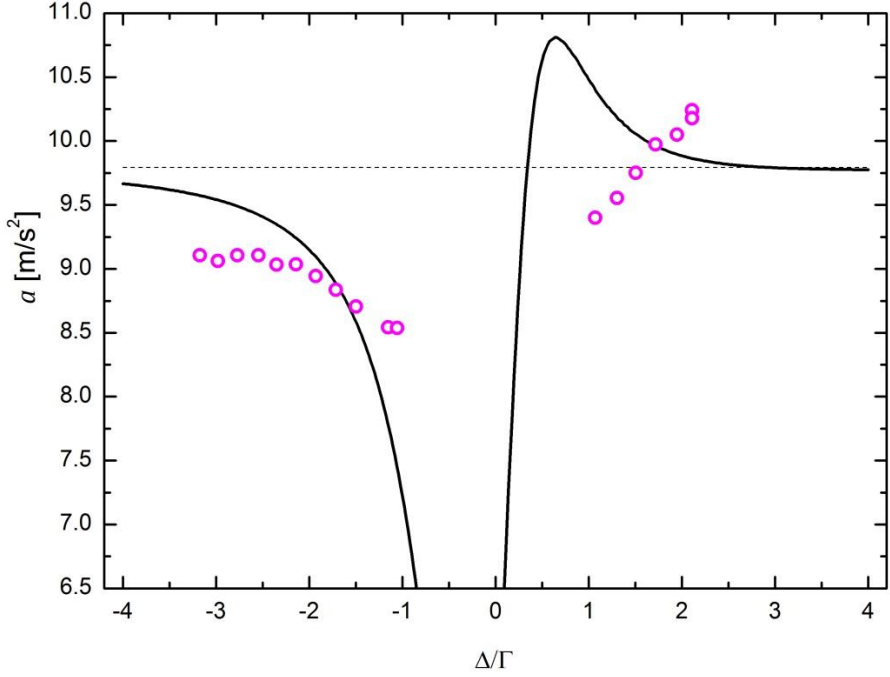

**Fig. S2. The calculated accelerations with the false assumption of**
$a_\phi'' = -4\pi G(\rho_c - \rho_{bg})\beta'^2 \sigma_p/2$. Here $\rho_c = \rho_{bg} + \delta E/c^2$ and $\delta E$ are described in Eq. (S5).
Circles are the experimental data. The solid curve is calculated by $a_{tot}'' = g - a_f - a_\phi''$, in which the temperature of the background atoms is $T \sim 300\,\text{K}$, the interaction time between the probe laser light and the atomic system is estimated by $t_d \sim \sqrt{2d/g}$, $\beta' = 8 \times 10^{16}$ and other parameters are the same as those in Fig. 2. The calculated curve cannot be adjusted to agree with the experimental dada. The false coupling $\beta' = 8 \times 10^{16}$ corresponding to the coupling parameter with the dimension of mass is $M' \equiv M_{Pl}/\beta' \approx 30\,\text{GeV}/c^2$. This value is much smaller than the lower bound of $M > 10^4\,\text{GeV}/c^2$ (*32*), which means that the assumption of $a_\phi'' = -4\pi G(\rho_c - \rho_{bg})\beta'^2 \sigma_p/2$ is false.